\documentclass{PoS}  

\title{Exploring the QCD phase diagram through high energy nuclear collisions: An overview}

\ShortTitle{QCD Phase Diagram}

\author{\speaker{Bedangadas Mohanty}\thanks{Supported by DST Swarna Jayanti Fellowship}\\
        School of Physical Sciences, National Institute of Science
        Education and Research, Bhubaneswar - 751005, India\\
        E-mail: \email{bedanga@niser.ac.in}}

\abstract{We present an overview of the status of the studies related to the
  phase diagram of strong interactions through high-energy nuclear collisions. 
  We discuss both the theoretical and experimental status
  of establishing the various QCD phase structures in the phase diagram, such as,
  quark-gluon phase, quark-hadron crossover, crossover temperature, 
  phase boundary and critical point.}

\FullConference{8th International Workshop on Critical Point and Onset of Deconfinement,\\
		March 11 to 15, 2013\\
		Napa, California, USA}

\begin{document}

\section{Introduction}
A phase diagram displays information on the structure of the matter as it manifests various degrees of 
freedom under different external conditions. 
One of the most widely studied phase diagram in science is that of water. In this case the underlying 
interaction is electromagnetic, one of the four basic interactions that occur in nature. 
The goal of high-energy nuclear collisions is to establish a similar phase diagram for a system whose 
underlying interaction is due to the strong force. The phase diagram of strong interaction on one hand 
displays the interplay of the chiral and center symmetry as a function of quark masses~\cite{Laermann:2003cv}, 
on the other hand can be represented as a graph which shows variation of temperature ($T$) versus the chemical potential ($\mu$) 
~\cite{Fukushima:2010bq} associated with conserved charges like baryon number ($B$), electric charge ($Q$) and strangeness number ($S$). 
The later is the one which can be experimentally studied. 

%
\begin{figure}
\begin{center}
\includegraphics[scale=0.4]{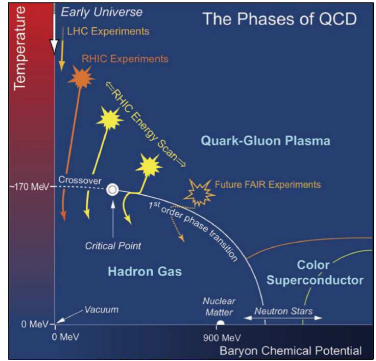}
\caption{\label{fig:pd}
(Color online) The conjectured phase diagram of strong interactions~\cite{phasedia}.}
\end{center}
\end{figure}

In the heavy-ion colliding systems, it is found that values of $\mu_{Q}$ and $\mu_{S}$ are small~\cite{Abelev:2008ab} and hence the 
phase diagram of strong interaction reduces to a two dimensional graph of $T$ vs. $\mu_{B}$ as shown in Fig.~\ref{fig:pd}. 
Further, the analysis of particle yields in the heavy-ion collisions and their comparison to statistical 
models suggests that $T$ and $\mu_{B}$  vary in opposite manner with center of mass energy ($\sqrt{s_{\rm NN}}$) 
at the chemical freeze-out~\cite{Cleymans:2005xv}. The  $\mu_{B}$ decreases with $\sqrt{s_{\rm NN}}$ while $T$ increases with increase 
in $\sqrt{s_{\rm NN}}$~\cite{BraunMunzinger:2007zz}. Thus changing the $\sqrt{s_{\rm NN}}$ one can vary the two axes of phase diagram, 
 $T$ and $\mu_{B}$, and  experimentally get access to a large part of the phase space. 
The Beam Energy Scan (BES) program has been designed based on this idea for the study of the phase structure of the 
quantum chromodynamic (QCD) phase diagram~\cite{Abelev:2009bw,Mohanty:2011nm}. 
The phase diagram in Fig.~\ref{fig:pd} shows the experimentally 
accessible parts using heavy-ion collisions at the Large Hadron Collider (LHC), Relativistic Heavy Ion Collider 
(RHIC) and through future experiments (like CBM at GSI and NICA at JINR). 
The phase diagram displays a rich phase structure, however the experimentally accessible part corresponds to 
 some of the following distinct structures: de-confined phase of quarks and gluons, hadronic phase, critical point, 
crossover line and a crossover at low $\mu_{B}$. We discuss in the sections below the progress towards 
establishing these phase structures, both theoretically and experimentally.

\section{Crossover and Crossover temperature}
%
\begin{figure}
\includegraphics[scale=0.4]{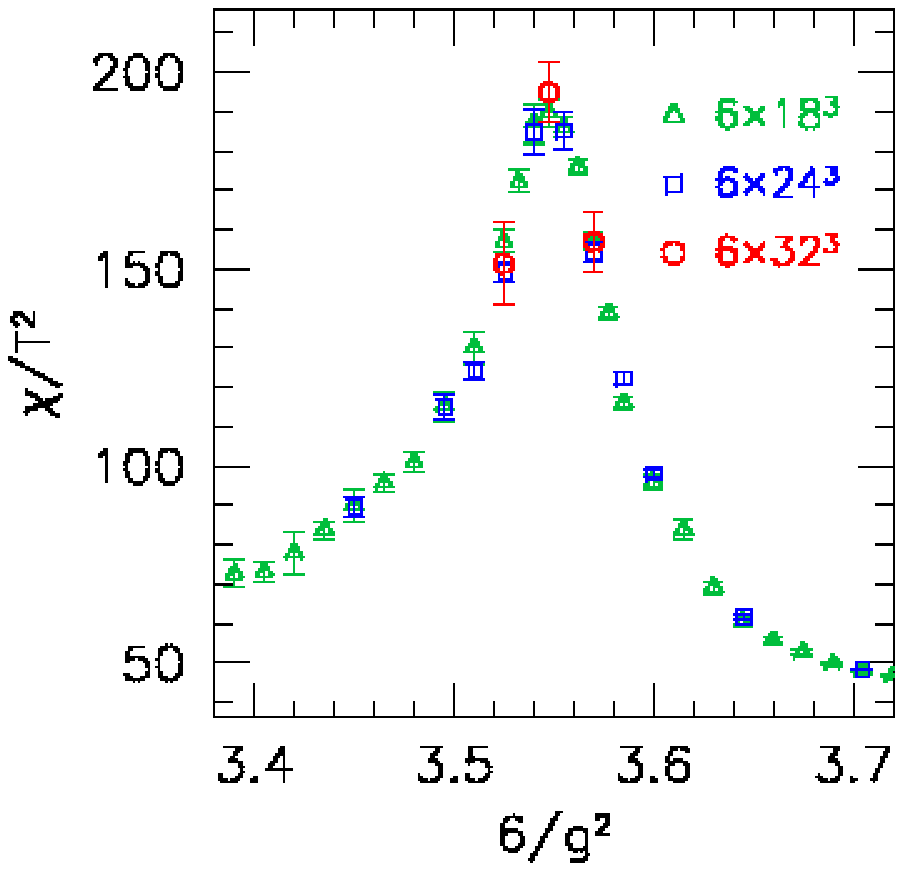}
\includegraphics[scale=0.4]{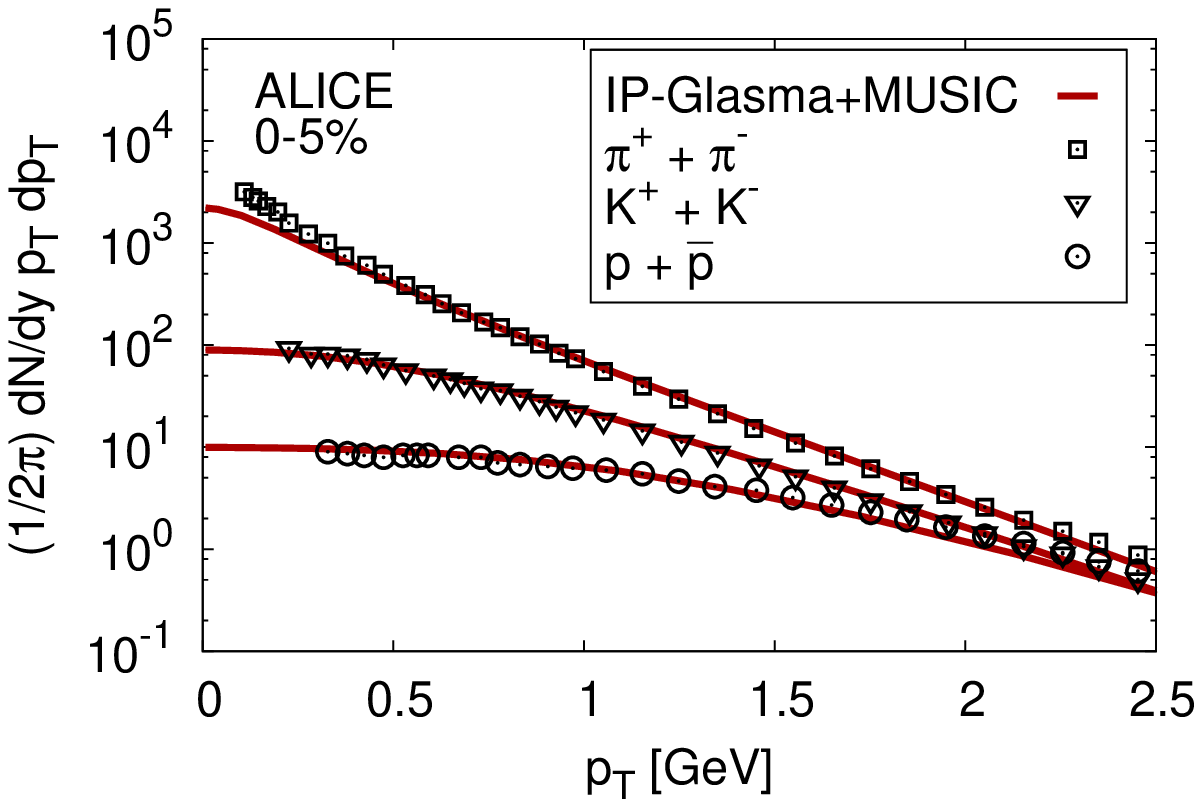}
\includegraphics[scale=0.4]{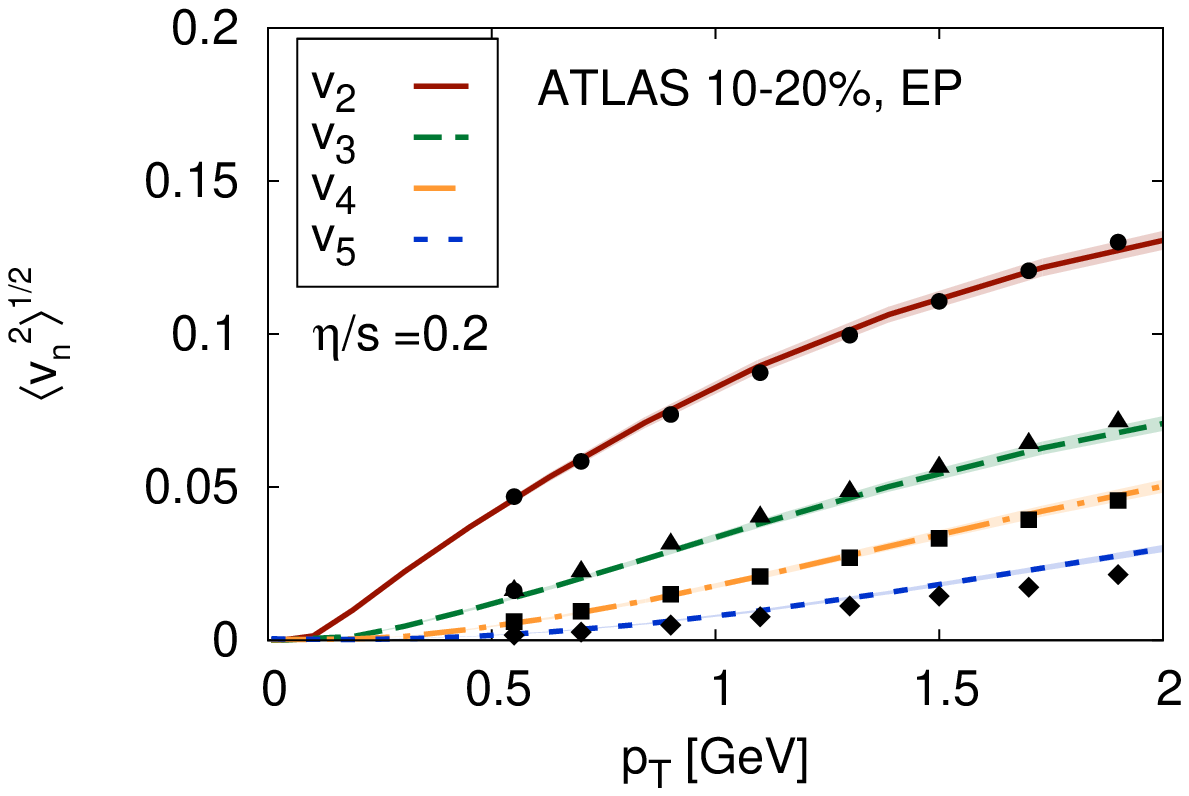}
\caption{\label{fig:co}
(Color online) Left panel: chiral susceptibility versus  6/$g^2$ ($g$ is the gauge coupling)~\cite{Aoki:2006we}.
Middle panel: Hydrodynamical calculation of identified hadron transverse momentum 
compared to experimental data from the ALICE collaboration, and Right panel: Root-mean-square
anisotropic flow co-efficients as a function of transverse momentum compared to experimental
data by the ATLAS collaboration~\cite{Gale:2012rq}.}
\end{figure}
\subsection{Crossover}
QCD calculations on lattice at high temperature and $\mu_{B}$ = 0 MeV have established the quark-hadron transition 
to be a crossover~\cite{Aoki:2006we}. Figure~\ref{fig:co} shows the lattice chiral susceptibility 
$\chi(N_s,N_t)$ = $\partial^2$/$(\partial$$m_{ud}^2)$($T/V$)$\cdot\log Z$, where $m_{ud}$ is the mass of the 
light u,d quarks, $N_s$ is the spatial extension, $N_\tau$ euclidean time extension,  and $V$ the system volume. 
The susceptibility plotted as a function of 6/$g^2$ ($g$ is the gauge coupling and $T$ grows with  6/$g^2$) 
shows a pronounced peak around the transition temperature ($T_c$). The peak and width are independent of volume (varied by a factor 8) 
thereby establishing the transition to be an analytic cross-over~\cite{Aoki:2006we}.
For a first-order phase transition the height of the susceptibility peak should have been $\propto$ $V$ and
the width of the peak $\propto$ 1/$V$, while for a second-order transition the
singular behaviour should have been $\propto$ $V^\alpha$, $\alpha$ is a 
critical exponent. Using the crossover equation of state for the quark-hadron transition
in a hydrodynamic based model, the experimental data on invariant yields of charged hadrons and 
various order azimuthal anisotropy as a function of transverse momentum at LHC are 
nicely explained (shown in Fig~\ref{fig:co})~\cite{Gale:2012rq}. Lending indirect support to the transition 
being a crossover at small $\mu_{B}$.

\subsection{Crossover temperature}
\begin{table}
\begin{center}
\caption{Crossover temperature at $\mu_{B}$ = 0 MeV along with the references.
\label{table1}}
\begin{tabular}{|l|c|c|c|c|c|r}
\hline
Observable&Transition temperature & Ref\\
\hline
$\chi_{\bar{\psi}\psi}$  & 151(3)(3) MeV  &~\cite{Aoki:2009zzc}\\
\hline
$\chi_{\bar{\psi}\psi}$  & 154(9) MeV     &~\cite{Bazavov:2011nk}\\
\hline
$\chi_s$                 & 175(2)(4) MeV  &~\cite{Aoki:2009zzc}\\
\hline
$L$                      & 176(3)(4) MeV  &~\cite{Aoki:2009zzc}\\
\hline
Baryon correlations      & 175(1)(7) MeV  &~\cite{Gupta:2011wh}\\
\hline
\end{tabular}
\end{center}
\end{table}

The point of sharpest change in temperature dependence of the
chiral susceptibility ($\chi_{\bar{\psi}\psi}$), the strange quark number susceptibility ($\chi_s$) 
and the renormalized Polyakov-loop ($L$) are used to estimate the crossover temperature in the lattice calculations. 
There is a clear agreement between various lattice QCD estimates of chiral crossover temperature 
using $\chi_{\bar{\psi}\psi}$~\cite{Aoki:2009zzc,Bazavov:2011nk}.
The observables ($\chi_s$ and $L$) that provide important insights into deconfining aspects of the crossover 
shows a slightly higher transition temperature. But with a width of around 15 MeV in temperature estimates, it is
difficult to make a concerte statement on the difference between deconfinement and chiral crossover temperatures. 
Moreover there are unresolved discussions on the establishment of Polyakov loop expectation and strange quark number 
susceptibilities to critical behaviour in the light quark mass regime~\cite{Bazavov:2011nk}. The crossover temperature 
situation is summarised in the table~\ref{table1}.

\section{Quark-Gluon Phase}

The results from heavy-ion collisions at relativistic high energies have clearly demonstrated 
the formation of a de-confined system of quarks and gluons at RHIC~\cite{Arsene:2004fa,Back:2004je,Adams:2005dq,Adcox:2004mh,Gyulassy:2004zy} and LHC~\cite{Singh:2013fha}.  The produced system 
exhibits copious production of strange hadrons~\cite{Abelev:2008zk}, shows substantial collectivity developed in the partonic phase~\cite{Abelev:2007rw,Afanasiev:2007tv}, 
exhibits suppression in high transverse momentum ($p_{T}$) hadron production relative to $p$+$p$ collisions~\cite{Adare:2008qa,Agakishiev:2011dc}, suppression in  quarkonia production relative to $p$+$p$ collisions~\cite{Chatrchyan:2012lxa},  and
small fluidity as reflected by a small value of viscosity to entropy density ratio ($\eta/s$)~\cite{Luzum:2008cw}. 
All these at temperatures and energy densities much larger than predicted by lattice QCD calculations 
for a quark-hadron transition. Some of these clear signatures are given in the table~\ref{table2}.

\begin{table}
\caption{Observables reflecting quark-gluon phase (QGP) formation in heavy-ion collisions at RHIC and LHC along with the references.
\label{table2}}
\begin{tabular}{|l|c|c|c|c|c|r}
\hline
Observable& Observation  & Remark & Reference\\
\hline
Nuclear modification factor  & $<$ 1       & at high $p_{T}$ for hadrons  &~\cite{Adler:2006hu,Aamodt:2010jd,CMS:2012aa}\\
\hline
Temperature                  & $>$ 300 MeV & direct photons                &~\cite{Adare:2008ab}\\
\hline
Strangeness enhancement      & $>$ 1       & crucial role by $\phi$ mesons &~\cite{Abelev:2008zk}\\
\hline
Constituent quark scaling   & for hadrons and nuclei & elliptic flow       &~\cite{Abelev:2007rw,Afanasiev:2007tv}\\
\hline
Dynamical charge correlations  & observed  & same and opposite charges &~\cite{Abelev:2009ac,Abelev:2012pa}\\
\hline
Quarkonia suppression          & observed  & $J/\Psi$ and $\Upsilon$ &~\cite{Chatrchyan:2012lxa,Adare:2006ns}\\
\hline
\end{tabular}
\end{table}

Here we only discuss one observable
called the nuclear modification factor ($R_{\rm {AA}}$). $R_{\rm {AA}}$ is defined as 
$\frac{dN_{AA}/d\eta d^2 p_{T}} {T_{AB} d\sigma_{NN}/d\eta d^2 p_{T}}$, here the overlap 
integral $T_{AB} = N_{binary}/\sigma_{inelastic}^{pp}$ with $N_{binary}$ being the
number of binary collisions commonly estimated from Glauber model calculation and $d\sigma_{NN}/d\eta d^2 p_{T}$
is the cross section of charged hadron production in $p$+$p$ collisions.  $R_{\rm {AA}}$ value of less than one is 
attributed to energy loss of partons in QGP and phenomenon is referred to as the jet quenching in a dense partonic
matter~\cite{Wang:1991xy}. It is one of the established signature of QGP formation in heavy-ion collisions. 

\begin{figure}
\includegraphics[scale=0.8]{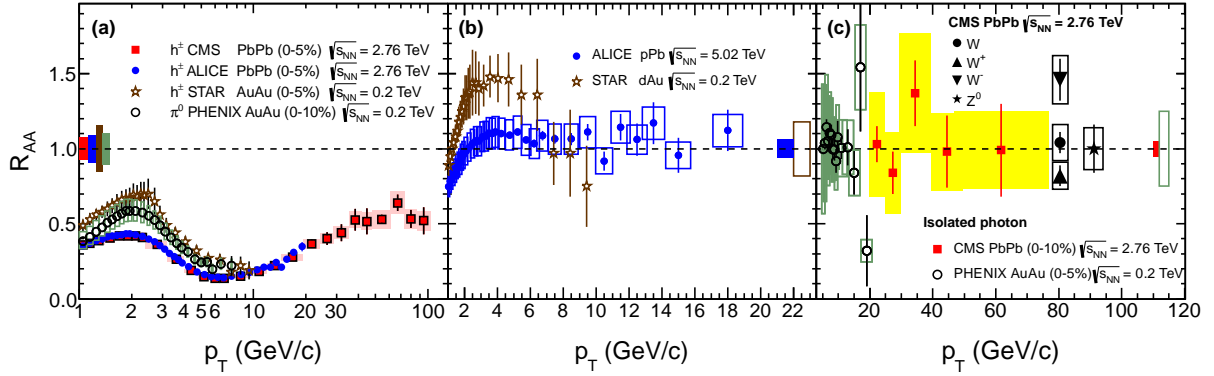}
    \caption{\label{fig:fig13}
     (Color online) (a) Nuclear modification factor $R_{\rm {AA}}$ of charged hadrons measured 
by ALICE~\cite{Aamodt:2010jd} and CMS~\cite{CMS:2012aa}  experiments at midrapidity.
Also shown the  $R_{\rm {AA}}$ of charged hadrons at midrapidity measured by STAR~\cite{Adams:2003im} 
and $R_{\rm {AA}}$ of $\pi^{0}$ at midrapidity measured by PHENIX~\cite{Adler:2006hu}.
(b) Comparison of nuclear modification factor for charged hadrons versus $p_{T}$ 
at midrapidity for minimum bias collisions in $d$+Au collisions at  
$\sqrt{s_{\mathrm {NN}}}$ = 200 GeV~\cite{Adams:2003im} and $p$+Pb collisions at  $\sqrt{s_{\mathrm {NN}}}$ = 5.02 TeV~\cite{ALICE:2012mj}. 
(c) The nuclear modification factor  versus $p_{T}$ for isolated photons in central 
nucleus-nucleus collisions at $\sqrt{s_{\mathrm {NN}}}$ = 200 GeV~\cite{Adler:2005ig} and 2.76 TeV~\cite{Chatrchyan:2012vq}. Also shown
are the $R_{\rm {AA}}$ of $W^{\pm}$~\cite{Chatrchyan:2012nt} and $Z$ bosons~\cite{Chatrchyan:2011ua} at LHC energies. 
The boxes around the data denote $p_{T}$-dependent systematic uncertainties. 
The systematic uncertainties on the normalisation are shown as boxes at $R_{\rm {AA}}$ = 1.}
  \end{figure}

Figure~\ref{fig:fig13} shows the  $R_{\rm {AA}}$ of various particles produced in heavy-ion collisions at RHIC and LHC.
In Fig.~\ref{fig:fig13}(a), we observe that the  values $R_{\rm {AA}}$ $<$ 1 and at RHIC are higher
compared to those at LHC energies up to  $p_{T}$ $<$ 8 GeV/$c$~\cite{Aamodt:2010jd,CMS:2012aa} . 
In Fig.~\ref{fig:fig13}(b), we observe that the nuclear modification factors for $d$+Au collisions at
$\sqrt{s_{\mathrm {NN}}}$~=~200 GeV~\cite{Adams:2003im} and  $p$+Pb collisions at  
$\sqrt{s_{\mathrm {NN}}}$ = 5.02 TeV~\cite{ALICE:2012mj} are greater 
than unity for the $p_{T}$ $>$ 2 GeV/$c$. The nuclear modification factor value in $p(d)$+A
collisions not being below unity strengthens the argument (from experimental point of view) that a hot and dense
medium of color charges is formed in A+A collisions at RHIC and LHC.  In Fig.~\ref{fig:fig13}(c), we show 
the $R_{\rm {AA}}$ of particles than do not participate in strong interactions and  some of them are most likely 
formed in the very early stages of the collisions. These particles (photon~\cite{Adler:2005ig,Chatrchyan:2012vq}, 
$W^{\pm}$~\cite{Chatrchyan:2012nt} and $Z$~\cite{Chatrchyan:2011ua}  bosons)
have a $R_{\rm {AA}}$ $\sim$ 1, indicating that the $R_{\rm {AA}}$ $<$
1, observed for hadrons in A+A collisions, are due to the strong 
interactions in a dense medium consisting of color charges. 

 %

\section{Crossover line}

 %
\begin{figure}
\includegraphics[scale=0.4]{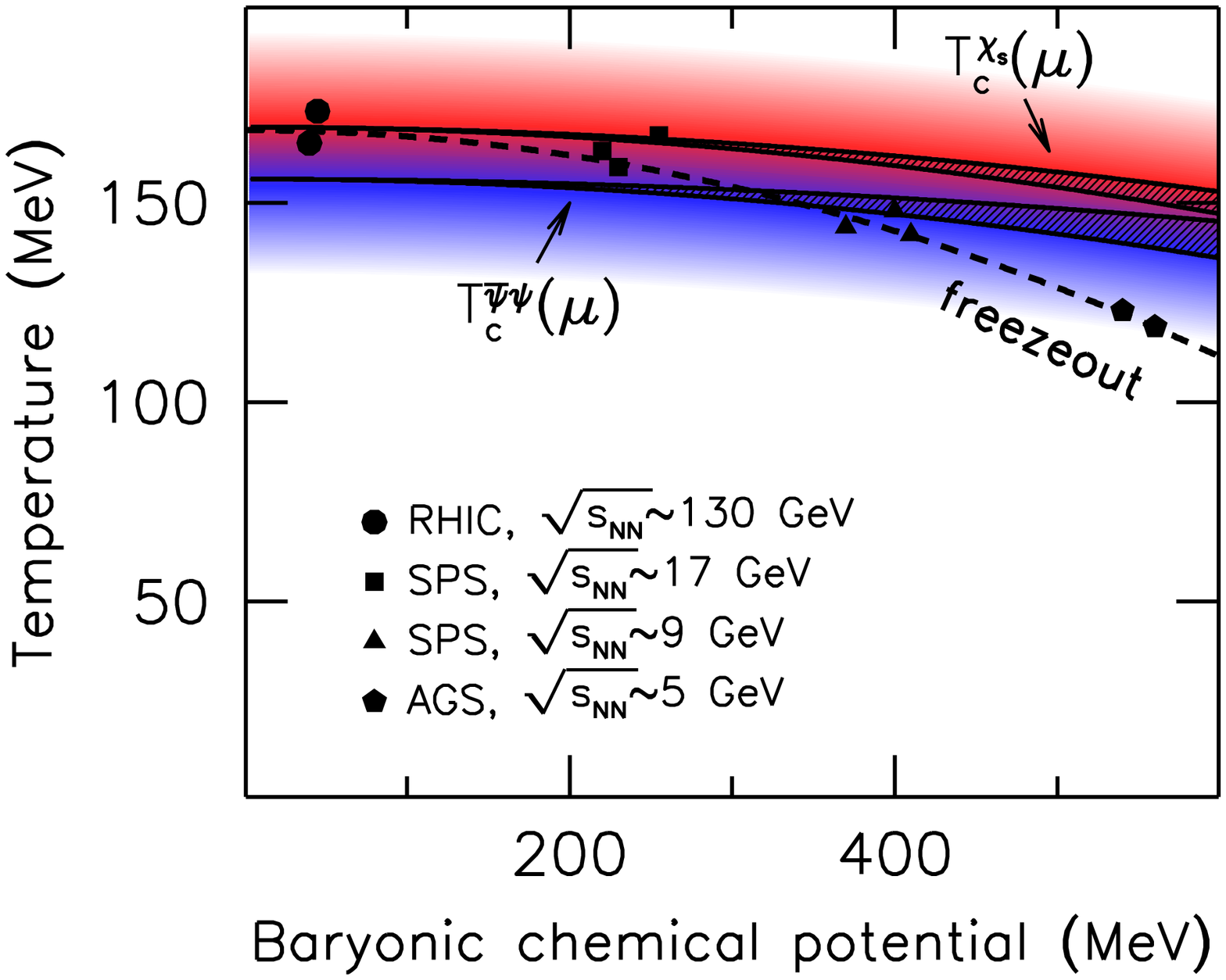}
\includegraphics[scale=0.45]{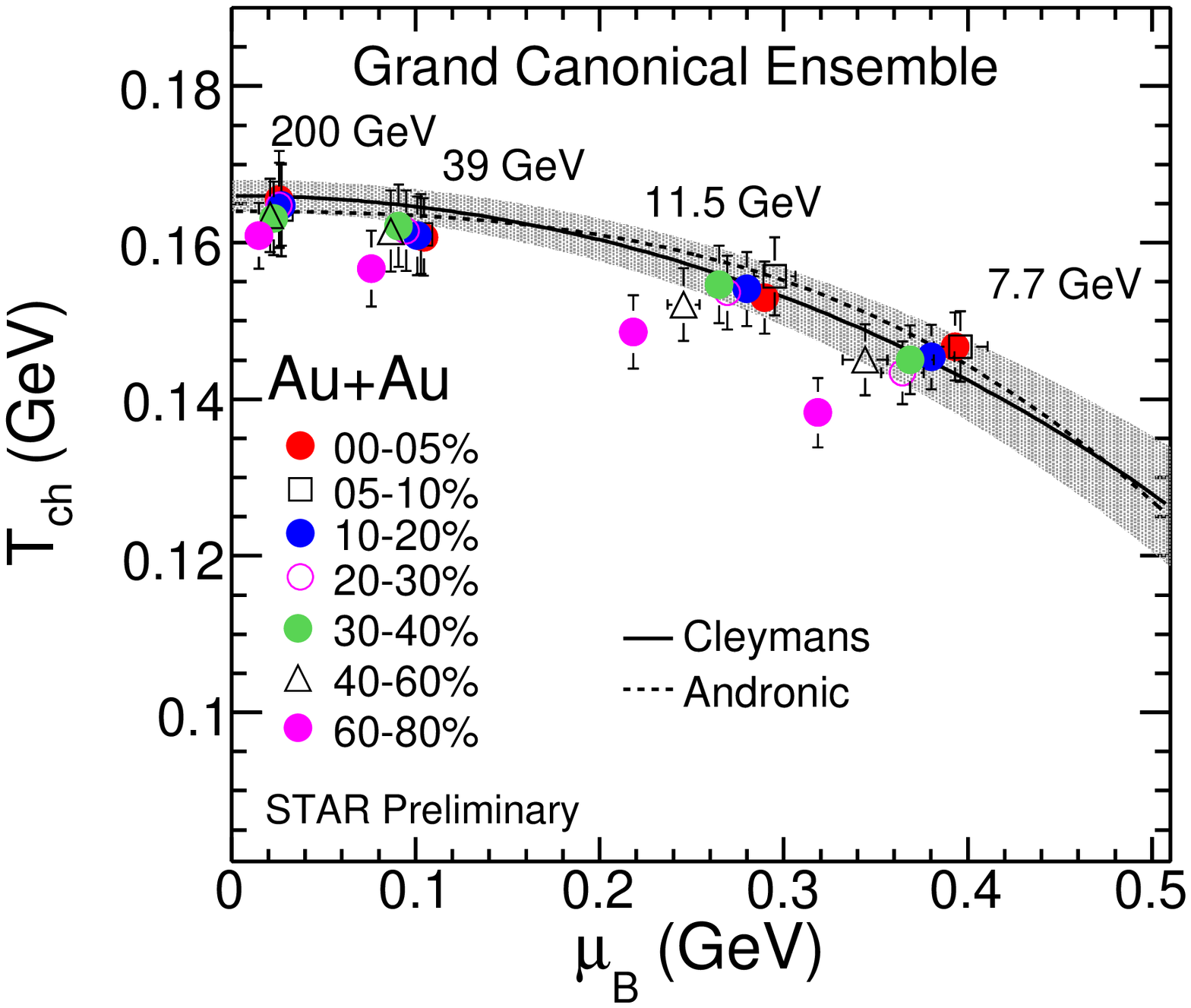}
\caption{\label{fig:tl}
(Color online) Left panel: The crossover line between the  quark-gluon and hadronic phases is 
represented by the coloured area (blue and red correspond to the transition
regions obtained from the chiral condensate and the strange susceptibility,
respectively)~\cite{Endrodi:2011gv}. Right panel: Variation of $T_{\rm {ch}}$ with $\mu_{B}$ for different energies 
and centralities. The curves represent the theoretical calculations. }
\end{figure}

\subsection{Theory estimates}
The quark-hadron transition at $\mu_{B}$ = 0 is a crossover~\cite{Aoki:2006we} and one of the important aspects of the phase diagram is to 
trace out the crossover temperature as we increase $\mu_{B}$. Besides the actual value of the curvature 
of the crossover line a particularly interesting question is whether the transition becomes weaker or 
stronger as $\mu_{B}$  grows (does it lead to a real phase boundary ?) and how close it is to the chemical freeze-out line. 
A recent lattice estimate is shown in Fig.~\ref{fig:tl}~\cite{Endrodi:2011gv}. Two crossover lines are defined with two quantities, 
the chiral condensate and the strange quark number susceptibility. The width of the bands represent the 
statistical uncertainty of $T_c(\mu)$ for the given $\mu$ coming from 
the error of the curvature for both observables. The dashed line 
is the freeze-out curve  from heavy ion experiments~\cite{Cleymans:1998fq}. It appears that the freeze-out
line is quite close to the transition line for a large range of values of  $\mu_{B}$. 
The right panel of Fig.~\ref{fig:tl} shows the estimates of chemical freeze-out temperature ($T_{\rm {ch}}$)
and  $\mu_{B}$ using a statistical model from the RHIC BES program~\cite{Kumar:2012fb}. One observes interesting dependence 
of $T_{\rm {ch}}$ vs. $\mu_{B}$ unfolding at lower beam energies.

\subsection{Turn-off of QGP signatures}

\begin{figure}
\includegraphics[scale=0.6]{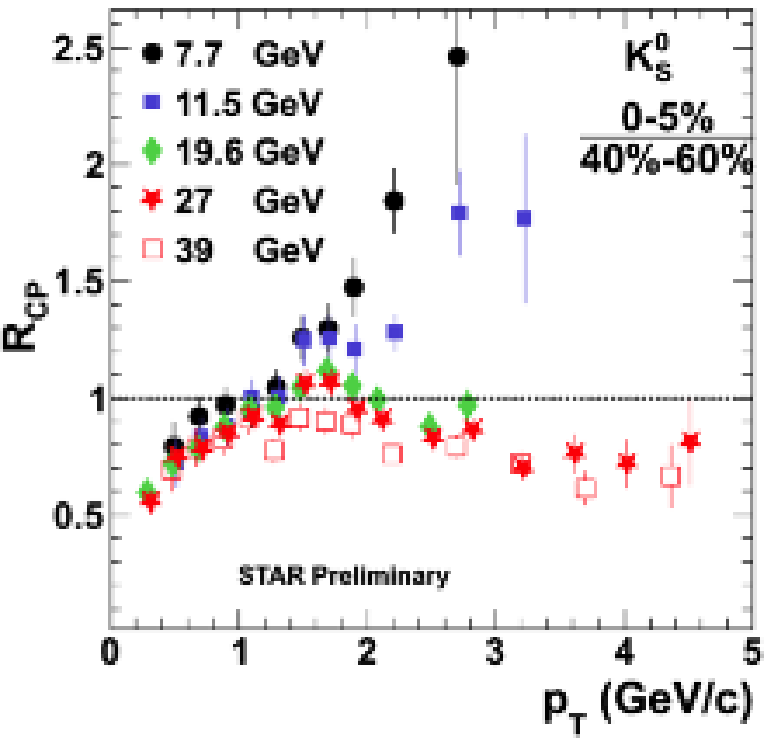}
\includegraphics[scale=0.45]{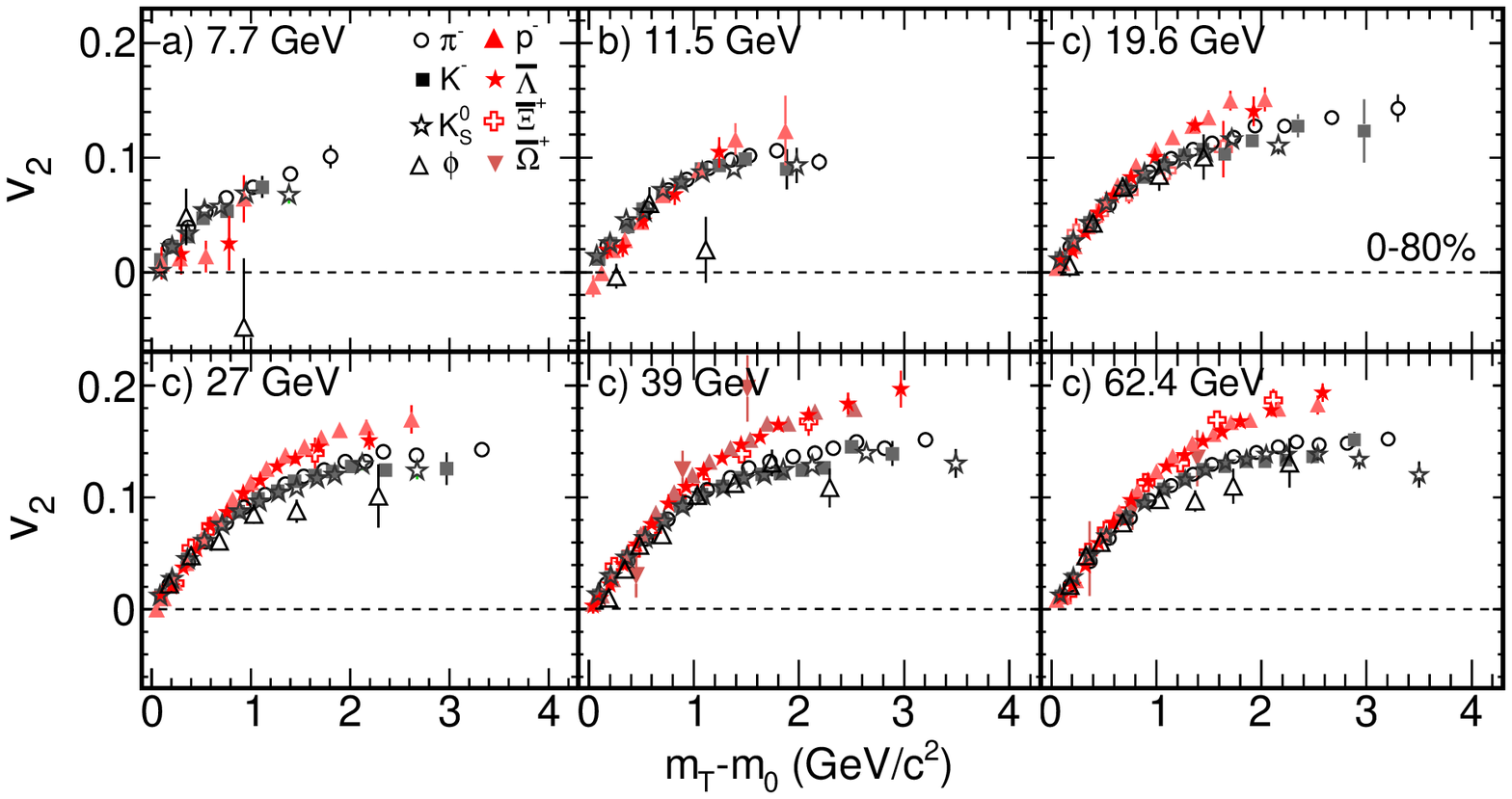}
\caption{\label{fig:toff}
(Color online) Left panel: Nuclear modification factor for $K^{0}_{S}$ mesons as a function
of transverse momentum for various  $\sqrt{s_{\mathrm {NN}}}$~\cite{Zhu:2012pg}. Right panel: Azimuthal 
anisotropy parameter, $v_{2}$ as a function of transverse mass ($m_{T}$) minus the mass
of the hadrons for various  $\sqrt{s_{\mathrm {NN}}}$~\cite{Adamczyk:2013gw}. }
  \end{figure}

An experimental way to demonstrate the quark-hadron transition is to show the turn-off of 
QGP signatures (like those discussed in section 3) as the collision energy is dialed down.
This interesting test is being carried out at the BES program in RHIC. Two such  
results are shown in Fig.~\ref{fig:toff}. For collision energies around 11.5 GeV the 
nuclear modification factor for $K^{0}_{S}$ mesons becomes $>$ 1 at $p_{T}$ $>$ 2 GeV/$c$, whereas
it gradually goes below unity for higher beam energies~\cite{Zhu:2012pg}. The baryon-meson splitting of azimuthal
anisotropy parameter $v_{2}$ (which is the basis for the claim of partonic collectivity 
at RHIC) reduces as the beam energy is dialed down and vanishes for  $\sqrt{s_{\mathrm {NN}}}$ = 11.5 and 7.7 GeV~\cite{Adamczyk:2013gw}.
These are an experimental indication that for  $\sqrt{s_{\mathrm {NN}}}$ = 11.5 GeV and below hadronic
interactions dominate as signatures associated with QGP phenomena seem smoothly getting turned-off. 
 %

\section{Critical Point}

\begin{figure}
\begin{center}
\includegraphics[scale=0.5]{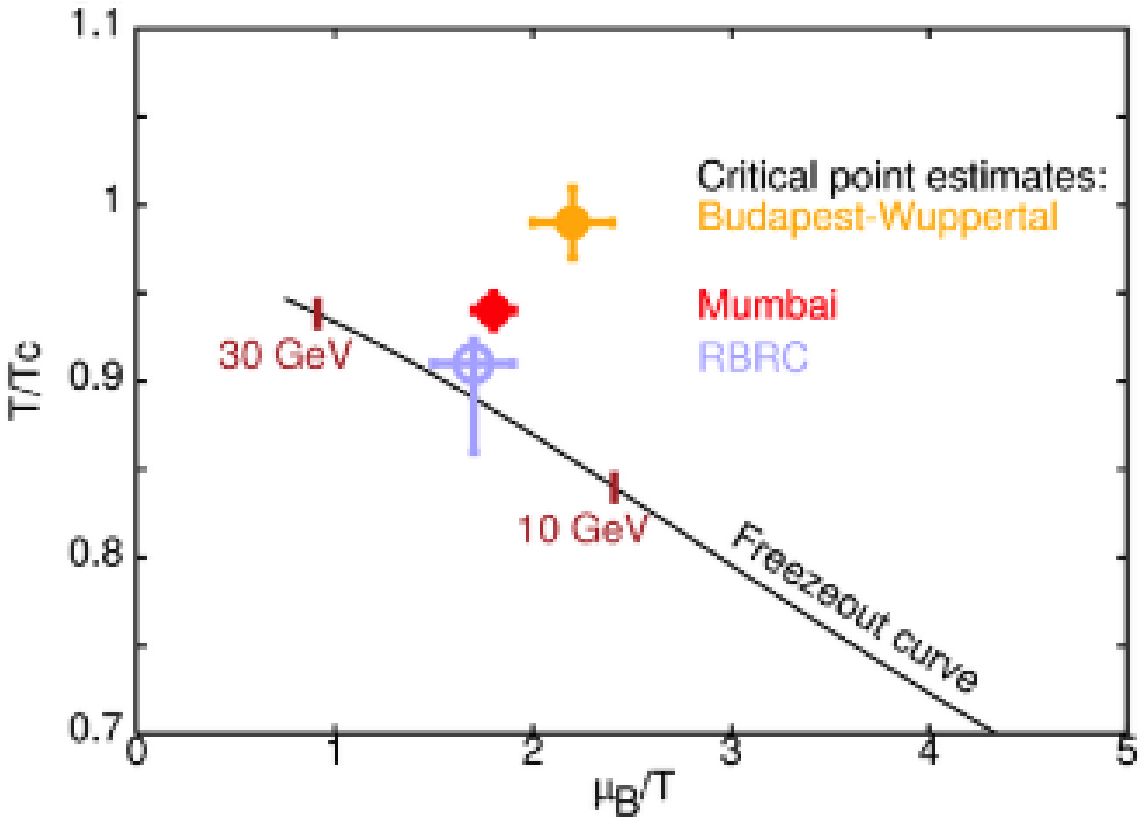}
\includegraphics[scale=0.5]{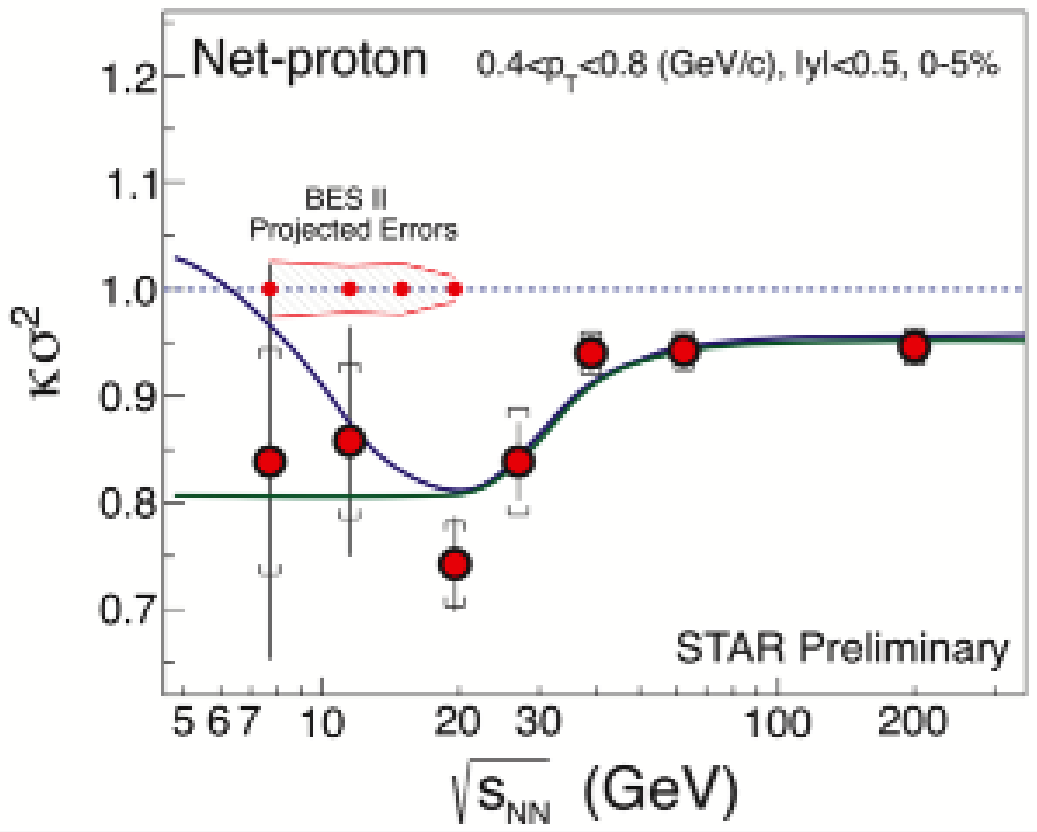}
\caption{\label{fig:cp}
(Color online) Left panel: Estimates of position of critical point from various lattice QCD calculations~\cite{Gupta:2010zzc,Gupta:2011zzd,Gupta:2012zze}. 
Right panel: $\kappa \sigma^2$ for net-proton distributions as a function of $\sqrt{s_{\mathrm {NN}}}$ in RHIC BES
program~\cite{Luo:2013saa}. Also shown are the projected statistical error in 2nd phase of BES program.}
\end{center}
  \end{figure}

Several QCD based models predict the existence of an end point or critical point (CP) at high $\mu_{B}$ 
for the first order phase transition in the QCD phase diagram. However the exact
location depends on the model assumptions used~\cite{Stephanov:2004wx}. Given the 
ambiguity in predictions of CP in models, studies on lattice was expected to provide reliable estimates~\cite{Gavai:2004sd}. 
However lattice calculations at finite $\mu_{B}$ are difficult due to {\it sign problem}.
There are several ways suggested to overcome this issue. (i) Reweighting the partition function
in the vicinity of transition temperature and $\mu$ = 0~\cite{Fodor:2004nz}, (ii) Taylor expansion of thermodynamic
observables in $\mu$/T about $\mu$ = 0~\cite{Gavai:2003mf}  and (iii) Choosing the chemical potential to be imaginary
will make the fermionic determinant positive~\cite{Philipsen:2009yg}.  The first two methodologies yield an 
existence of CP, whereas the third procedure gives a CP only when the first co-efficient in the 
Taylor expansion of generic quark mass on the chiral critical surface ($m_{c}$) as a function of  
$\mu$/T ($\frac{m_c(\mu)}{m_c(0)} = 1+\sum_{k=1} c_{k} \left(\frac{\mu}{\pi T_c}\right)^{2k}$) 
is positive. The lattice calculations which yield a CP on phase diagram are shown in Fig.~\ref{fig:cp}~\cite{Gupta:2010zzc,Gupta:2011zzd,Gupta:2012zze}.
However these calculations have to overcome some of the common lattice artifacts like, lattice spacing,
physical quark masses, volume effect and continuum limit extrapolation.

 %

In the experimental side, the characteristic signature of CP is large fluctuations in event-by-event
conserved quantities like net-charge, net-baryon number and net-strangeness. The
variance of these distributions ($\langle (\delta N)^{2} \rangle$) are proportional to square
of the correlation length ($\xi$). The finite size and finite time effects attained 
in high energy heavy-ion collisions, limits the value of the $\xi$ 
achieved in the collisions, thereby making it extremely challenging to measure 
in the experiments. Motivated by the fact that non-Gaussian features in above observables 
increase if the system freezes-out closer to QCP, it has
been suggested to measure higher moments (non-zero skewness and kurtosis 
indicates non-Gaussianity) of net-charge or net-baryon number distributions. 
Further it has been shown that higher moments ($\langle (\delta N)^{3} \rangle$ 
$\sim$ $\xi^{4.5}$ and $\langle (\delta N)^{4} \rangle$ $\sim$ $\xi^{7}$)
have stronger dependence on $\xi$ compared to variance and hence have higher sensitivity~\cite{Stephanov:2008qz,Stephanov:2011pb,Asakawa:2009aj}.
In addition the moments are related to  susceptibilities~\cite{Cheng:2008zh}.
Motivated by all these, experiments are studying the variable $\kappa$$\sigma^2$ of net-proton 
distributions (a proxy for net-baryon),  to search for the CP. 
The $\kappa$$\sigma^2$ will be constant as per the CLT and have monotonic dependence
with $\sqrt{s_{\mathrm {NN}}}$ for non-CP scenarios. However as it is related to 
the ratio of baryon number susceptibilities in QCD models: 
$\kappa$$\sigma^2$ = $\frac{\chi^{(4)}_{\mathrm B}}{\chi^{(2)}_{\mathrm B}/T^2}$~\cite{Gavai:2010zn},
close to  CP it is expected to show a non-monotonic dependence on  $\sqrt{s_{\mathrm {NN}}}$.
Preliminary experimental results on  $\kappa$$\sigma^2$ value for net-proton distributions
measured in RHIC BES program is shown in right panel of Fig.~\ref{fig:cp}~\cite{Luo:2013saa}. Interesting trends
are observed indicating CP if exists in the phase diagram, have to be below 
$\sqrt{s_{\mathrm {NN}}}$ = 39 GeV~\cite{Aggarwal:2010wy}.

\section{Summary}

Significant progress has been made towards establishing the QCD phase diagram.  
From the QCD calculations on lattice it is now established theoretically that the quark-hadron 
transition at $\mu_{B}$ = 0 MeV is a crossover. This is indirectly supported by the experimental 
data, as models using the lattice based equation-of-state explain various measurements at RHIC and LHC.  
Lattice QCD calculations are in agreement that the chiral crossover temperature is around 154 MeV.
Other observables of quark-hadron crossover give a slightly higher values of crossover temperature
with large uncertainties.

High energy heavy-ion collision experiments have seen distinct signatures which suggest that the 
relevant degrees of freedom at top RHIC and LHC energies in the initial stages of the collisions are quark and gluons
and the system quickly approaches thermalization. The underlying mechanism for the fast thermalization is currently under 
study. Dialling down the beam energies to 11.5 GeV and below leads to a smooth turning-off of the QGP signatures, indicating
that hadronic interactions dominate. 
These observations in turn further support the formation of partonic matter at higher energy collisions.
Lattice QCD calculations of the crossover line indicates that they are
close to freeze-out line for a substantial part of the phase diagram. The experimental measurements of
freeze-out parameters in RHIC BES program reveals interesting temperature versus baryonic chemical potential
dependences at lower beam energies. 

Most calculations on lattice continue to indicate the possible existence of critical point for $\mu_{B}$ $>$ 160 MeV, 
this possibility have not been ruled out from the data at RHIC. The exact location is not yet known unambiguously. 
The experimental measurements though encouraging are inconclusive.

New phase structures are being proposed, like the existence of a 
quarkyonic phase around $\mu_{B}$ values corresponding to FAIR energies~\cite{McLerran:2009ve}. 
This is in addition to the confined and de-confined phases. The matter in such a phase is 
expected to have energy density and pressure that of a gas of quarks, and  yet is a chirally symmetric confined matter. 
Baryon-Baryon correlations to look for nucleation of baryon rich bubbles surrounded 
by baryon free regions could be a signature of such a phase. A summary of the status
of QCD phase diagram studies in theory and experiments is given in Table~\ref{table4}. \\

\begin{table}
\caption{QCD phase structure: Theory and Experiment status.
\label{table4}}
\begin{tabular}{|l|c|c|c|c|c|r}
\hline
Phase structure&Theory&Experiment& Remark\\
\hline
De-confined phase           & Yes        & Yes         & RHIC and LHC \\
\hline
Cross over                  & Yes        & Indirectly  & $\mu_{B}$ $\sim$ 0 MeV \\
\hline
Crossover temperature      & Yes        & Yes  & $\mu_{B}$ $\sim$ 0 MeV \\
\hline
Crossover line             & Uncertain  & Indications  & RHIC BES\\
\hline
Critical Point              & Uncertain  & Inconclusive  & $\mu_{B}$ $>$ 200 MeV \\
\hline
QGP properties              & Progress   & Progress      & Temperature, density, $\eta/s$\\
\hline
Hadronic phase properties   & Progress   & Progress      &  Freeze-out dynamics\\
\hline
New phases                  & Proposed   & Lack of proper observables  &  Quarkyonic phase\\
\hline
\end{tabular}
\end{table}

Acknowledgement: We thank Sourendu Gupta, Lokesh Kumar and Nu Xu for careful reading of the manuscript.
Financial support is obtained from the DST Swarna Jayanti Fellowship.

\end{document}